\documentclass[lettersize,journal]{IEEEtran}
\usepackage{amsmath,amsfonts}

\usepackage{array}
\usepackage[caption=false,font=normalsize,labelfont=sf,textfont=sf]{subfig}
\usepackage{textcomp}
\usepackage{stfloats}
\usepackage{url}
\usepackage{verbatim}
\usepackage{graphicx}
\usepackage{cite}

\usepackage{makecell}
\usepackage{multirow}
\usepackage{booktabs} 
\usepackage{longtable}
\usepackage{stfloats}  

\usepackage{algorithmic}
\usepackage{algorithm}
\usepackage{amssymb}
\usepackage{color} 

\begin{document}

\title{DKDNet: Dual Knowledge and Data-Driven Network for \\ Cross-Domain Automatic Modulation Classification}

\author{
	Shuang Wang, ~\IEEEmembership{Senior Member,~IEEE,} 
	Chenxu Wang,  
	Hantong Xing, ~\IEEEmembership{Student Member,~IEEE,}  \\
	Hanlin Mo, 	~\IEEEmembership{Member,~IEEE,}
	Lirong Han,
	 and Licheng Jiao, ~\IEEEmembership{Life Fellow,~IEEE}

\thanks{
	This work was supported in part by the National Natural Science Foundation of China under Grant 62271377; in part by the Key	Research and Development Program of Shaanxi Program	under Grant 2024GX-ZDCYL-02-08.	
	 (\textit{Corresponding author: Shuang Wang})}

\thanks{Shuang Wang, Chenxu Wang, Lirong Han, and Licheng Jiao are with the School of Artificial Intelligence, Xidian University, Xi’an 710071, China. (email: shwang@mail.xidian.edu.cn; wcxlocal@163.com;  hanlirong@xidian.edu.cn; lchjiao@mail.xidian.edu.cn).
Hantong Xing is with the School of Electronics And Information, Northwestern Polytechnical University, Xi’an 710072, China (email: xinghantong@126.com).
Hanlin Mo is with the Unmanned System Research Institute, Northwestern Polytechnical University, Xi’an 710072, China (e-mail: mohanlin@nwpu.edu.cn). 

}

}

\markboth{Journal of \LaTeX\ Class Files,~Vol.~14, No.~8, June~2026}%
{Shell \MakeLowercase{\textit{et al.}}: DKDNet: Dual Knowledge and Data-Driven Network for Cross-Domain Automatic Modulation Classification}


\maketitle

\begin{abstract}
The dynamics of communication environments induce significant distribution shifts across domains, challenging the generalization of deep learning-based automatic modulation classification (AMC) models. 
While existing UDA methods alleviate this problem by aligning source and target features, they give limited consideration to modulation-specific structures that remain informative across domain conditions.
In this paper, we consider signal prior knowledge, grounded in communication protocols and physical principles, as a potential way to enhance cross-domain representation learning. Given that different priors may vary in modulation discriminability, domain stability, and complementarity, this paper first analyzes five commonly adopted signal representations that instantiate different signal priors. From them, in-phase/quadrature (IQ), amplitude--phase (AP), and autocorrelation function (ACF) are selected as compact prior-guided inputs.
Based on that, a dual knowledge and data-driven network (DKDNet) is proposed for cross-domain AMC. The multi-representation feature encoder (MRFE) and dynamic lightweight fusion unit (DLFU) are designed to achieve unified representation learning and adaptive feature fusion, and the resulting fused features are optimized with modulation classification and adversarial domain alignment objectives.
Experiments on both simulated and public datasets validate the rationality of the prior selection and demonstrate the superiority of the proposed method. 

\end{abstract}


\section{INTRODUCTION}\label{sec:intro}

\IEEEPARstart{A}{utomatic} modulation classification (AMC) aims to identify the modulation type of received signals in a non-cooperative manner. As a key technique for radio-environment perception, AMC has been applied to both military and civilian communication scenarios, including cognitive radio~\cite{10818854,dong2024edge}, threat assessment~\cite{xu2020spatiotemporal,perenda2021learning}, and spectrum monitoring~\cite{rajendran2018deep,ke2021real}.

Traditional AMC methods mainly include likelihood-based and feature-based approaches~\cite{dobre2007survey}.
Likelihood-based methods formulate AMC as composite hypothesis testing using the conditional probability density of received waveforms, achieving statistical optimality but with high computational complexity~\cite{7931606,zhang2017cooperative}.
Feature-based methods classify signals using carefully designed hand-crafted features that provide compact and discriminative representations~\cite{yan2024automatic,huang2024generalized}.
Common features include instantaneous amplitude and phase (AP) \cite{muller2011front},
 constellation diagrams (CD) \cite{kumar2020automatic}, 
discrete Fourier transform (DFT)\cite{yu2003m}, 
and higher-order cumulants \cite{11153513,aslam2012automatic}. 
Each of them captures essential signal characteristics, such as frequency information\cite{kulin2018end}, the order of digital modulation\cite{peng2018modulation}, and periodicity\cite{gardner2002exploitation}.

Recently, deep learning (DL)-based AMC methods have emerged as a promising alternative to traditional approaches, as they leverage data-driven strategies to learn discriminative features directly from raw in-phase and quadrature (IQ) signals~\cite{o2016convolutional,10752883}.
However, such data-driven models usually assume consistent training and testing distributions, which is often violated in practical communication systems.
Factors such as transmission loss~\cite{goldsmith2005wireless}, multipath fading~\cite{SigDA}, environmental noise~\cite{10857965}, and RF-device calibration errors~\cite{rajendran2018deep} can induce significant distribution shifts.
These shifts often lead to severe performance degradation when DL-based models are deployed in unseen scenarios, challenging their practical application in real-world communication systems\cite{perenda2021learning,zhang2024open}.

To address the performance degradation caused by distribution shifts, researchers have adopted unsupervised domain adaptation (UDA) methods\cite{li2019deep, wang2025survey}. 
In UDA-based AMC, labeled source-domain samples and unlabeled target-domain samples from shifted distributions are jointly used to learn transferable features.

By introducing adversarial learning strategies, existing UDA methods usually map source and target samples into a shared feature space to mitigate performance degradation caused by domain shift. However, in wireless signal scenarios, domain shifts are often induced by intertwined channel variations, noise perturbations, and hardware discrepancies, making cross-domain AMC subject to more complex distribution changes\cite{SigDA,perenda2021learning}. Relying solely on data-driven feature alignment may therefore be insufficient to explicitly preserve modulation-specific structures that remain discriminative across domains.

Existing knowledge-embedded AMC studies have shown that signal priors can improve representation learning and recognition performance under fixed-channel conditions. Motivated by this observation, this work further investigates whether such priors can facilitate cross-domain representation learning.
These priors are usually derived from the physical principles and protocol structures of digital modulation~\cite{haykin1988digital,goldsmith2005wireless}, and are commonly introduced through explicit signal representations or auxiliary knowledge, such as CD~\cite{mao2021attentive}, AP~\cite{10769507}, statistical features~\cite{zheng2023toward}, and semantic information~\cite{ding2023data}.
Different from existing studies that mainly focus on classification accuracy or interpretability under fixed-channel conditions, this work explores whether signal priors can provide stable and modulation-discriminative structural cues for cross-domain representation learning.

In communication theory and signal processing, different signal representations characterize complementary properties of modulated signals and provide concrete forms for signal priors~\cite{peng2021survey,zhang2025revolution}.
Accordingly, we consider five commonly used representations, including IQ, AP, ACF, CD, and DFT, as candidates for prior-guided representation learning.
They describe the received signal from different perspectives: IQ preserves the raw complex waveform~\cite{o2016convolutional}; AP describes amplitude--phase evolution~\cite{meng2018automatic}; ACF captures statistical dependence and periodicity~\cite{gardner1989statistical}; CD reveals constellation geometry~\cite{peng2018modulation}; and DFT characterizes spectral structure~\cite{10042021}.
Nevertheless, a richer representation does not necessarily lead to better cross-domain generalization. Some representations may retain strong modulation-discriminative cues together with domain-specific distortions, whereas others may be more stable but less discriminative. Therefore, the key is to select and integrate signal priors that balance discriminability, complementarity, and cross-domain stability.

Motivated by this observation, we propose DKDNet, a dual knowledge and data-driven framework for cross-domain AMC. 
We first analyze five commonly adopted signal representations from the perspectives of modulation discriminability, domain discrepancy, and complementarity. 
Based on both theoretical analysis and empirical evidence, IQ, AP, and ACF are selected as the prior-guided inputs of DKDNet: IQ preserves detailed waveform information, AP provides explicit amplitude--phase dynamics, and ACF introduces relatively stable correlation-domain statistics. 
To effectively exploit these heterogeneous inputs, DKDNet employs a multi-representation feature encoder (MRFE) to extract representation-specific features in a unified latent space, and a dynamic lightweight fusion unit (DLFU) to adaptively fuse complementary information. 
The fused features are further optimized with modulation classification and domain-adversarial objectives, enabling the model to learn features that are both discriminative and transferable.
This structure is highly flexible, allowing a variety of AMC classifier and UDA methods to be jointly applied to the fused features for modulation recognition and cross-domain alignment, respectively.
Furthermore, we generated synthetic datasets, RML2025 Series, to evaluate the performance of AMC models under cross-domain conditions and have made the datasets publicly available. 
Experimental results show that DKDNet effectively improves the cross-domain generalization ability of AMC models, demonstrating the benefits of combining signal knowledge with data-driven learning approaches. 
\footnote{Code is available at: \url{https://github.com/FireTracer/DKDNet-AMC}.}

The main contributions of this paper are summarized as follows.
\begin{itemize}
	\item[$\bullet$]		
	The role of signal prior knowledge in cross-domain AMC is systematically investigated. 
	Five signal representations are analyzed from the perspectives of discriminability, domain discrepancy, and complementarity, providing a basis for selecting IQ, AP, and ACF as compact and transferable prior-guided inputs.

	\item[$\bullet$]
	DKDNet is proposed as a dual knowledge and data-driven framework for cross-domain AMC. Structural signal priors are introduced as representation-level inputs and integrated with data-driven feature learning to improve target-domain generalization under UDA.

	\item[$\bullet$]
	MRFE and DLFU are designed to encode heterogeneous signal representations in a unified feature space and adaptively fuse their complementary information. The resulting framework is compatible with different AMC backbones and UDA objectives.	
	
	\item[$\bullet$]	

	The RML2025 Series datasets are constructed with progressively intensified channel impairments, and extensive experiments are conducted to validate DKDNet. Consistent improvements are demonstrated in cross-domain performance, robustness, flexibility, and sample efficiency.

\end{itemize}

The remainder of this paper is organized as follows. 
Section II reviews related work on AMC and UDA-based AMC. 
Section III presents the problem formulation, signal-prior analysis, and DKDNet. 
Section IV describes the RML2025 Series datasets and reports experimental results. 
Section V concludes this paper.

\section{RELATED WORK}
This section reviews recent AMC methods in two categories: in-domain and cross-domain AMC, according to whether distribution shifts exist between training and testing.

\subsection{In-domain AMC}
In-domain AMC relies on the assumption that the training and test data are independent and identically distributed. 
Here, we broadly categorize existing approaches into three types: traditional methods, data-driven DL-based methods, and knowledge-embedded DL-based methods.

\textit{Traditional methods: }
Traditional AMC methods mainly include likelihood-based and feature-based approaches\cite{dobre2007survey}. 
Likelihood-based methods formulate AMC as composite hypothesis testing based on statistical criteria, achieving statistical optimality but suffering from high computational complexity~\cite{siran1,7931606}. 
Feature-based methods extract hand-crafted signal features and combine them with classifiers such as decision trees or support vector machines (SVMs)~\cite{muller2011front,kumar2020automatic}. 
Although these methods offer flexibility and robustness, their performance largely depends on feature design and prior signal-processing expertise.

\textit{Data-driven DL-based methods: }
Data-driven DL-based AMC methods learn discriminative features directly from raw signals, improving flexibility and classification performance. 
O'Shea et al. introduced a public AMC dataset and a pioneering convolutional neural network (CNN) for modulation classification~\cite{o2016convolutional}. 
Liu et al. compared several deep neural network (DNN) architectures and identified convolutional long short-term memory (LSTM) networks as an effective structure~\cite{liu2017deep}. 
Xing et al. proposed to transform feature learning into relationship learning through paired samples, thereby reducing sensitivity to distributional differences\cite{10752883}.

\textit{Knowledge-embedded DL-based methods: }
Recent studies have incorporated signal priors into DL-based AMC models to improve data efficiency, interpretability, and classification performance. 
Typical priors include CD~\cite{mao2021attentive}, AP information~\cite{chang2021multitask}, statistical hand-crafted features~\cite{zheng2023toward}, and modulation semantics~\cite{ding2023data}. 
Other studies further combine representations, such as CD with IQ~\cite{wang2019data}, time-frequency images with instantaneous autocorrelation maps~\cite{wang2019automatic}.

Different from these methods, this work investigates whether signal priors can provide modulation-discriminative and domain-stable cues for cross-domain representation learning.
By evaluating the discriminability, domain discrepancy, and complementarity of candidate representations, we select more transferable priors to enhance UDA generalization.

\subsection{Cross-domain AMC}
In practical communication systems, distribution discrepancies across domains are widespread, making cross-domain AMC increasingly important.
Perenda et al. showed that unknown channel parameters can significantly degrade AMC performance across different channel conditions~\cite{perenda2021learning}.
To mitigate this problem, adversarial UDA has been introduced to align source and target signal distributions~\cite{li2019deep,li2021domain}.
Bu et al. further studied adversarial transfer learning under varying sampling rates~\cite{bu2020adversarial}.
Recent studies have also considered more complex domain shifts caused by multiple signal impairments~\cite{SigDA} and label imbalance in cross-domain AMC~\cite{mei2024imbalanced}.

Collectively, these studies demonstrate the effectiveness of UDA in improving AMC generalization.
Nevertheless, most existing methods mainly rely on data-driven feature alignment, with limited consideration of modulation-specific structures that remain informative across domain conditions.
This work introduces signal priors into cross-domain AMC and proposes DKDNet to enhance domain generalization.

\section{PROPOSED METHOD}
This section formulates the cross-domain AMC problem, analyzes representative signal priors, and presents DKDNet, including MRFE, DLFU, and the task-specific heads for modulation classification and domain alignment.

\subsection{Problem Definition}
\label{sec:problem_definition}

\subsubsection{Signal Model}
\label{sec:signal_model}

In practical communication systems, the received signal is affected by the transmitter, the propagation channel, and the receiver. 
For a transmitted baseband signal $s(t)$, the received signal in domain $d$ can be modeled as
\begin{equation}
	x_d(t)
	=
	a_d e^{j(\phi_d-2\pi \Delta f_d t)}
	\big[s(t-\Delta t_d)\circledast H_d(t,\tau)\big]
	+
	n_d(t),
	\label{eq:signal_model}
\end{equation}
where $a_d$ denotes the amplitude scaling factor, $\Delta t_d$ is the timing offset, $\Delta f_d$ is the carrier frequency offset (CFO), $\phi_d$ is the phase offset, $H_d(t,\tau)$ denotes the domain-dependent multipath fading channel with delay component $\tau$, and $n_d(t)$ is the AWGN. 
The convolution operation $\circledast$ characterizes the interaction between the transmitted signal and the channel response. 
The variations of $a_d$, $\Delta t_d$, $\Delta f_d$, $\phi_d$, $H_d(t,\tau)$, and $n_d(t)$ across domains are the main sources of distribution shift in cross-domain AMC.

After sampling $x_d(t)$ with sampling period $T_s$, we obtain the discrete complex sequence $X_d[n]=x_d(nT_s)$, $n=1,\ldots,L$. 
Each complex sample can be decomposed into its in-phase and quadrature components as $X_d[n]=I_d[n]+jQ_d[n]$. 
For a signal segment with length $L$, the raw IQ representation is written as
\begin{equation}
	{X}_{\mathrm{IQ}}^d
	=
	\begin{bmatrix}
		\mathrm{Re}\{X_d[1]\}, \ldots, \mathrm{Re}\{X_d[L]\} \\
		\mathrm{Im}\{X_d[1]\}, \ldots, \mathrm{Im}\{X_d[L]\}
	\end{bmatrix}
	\in \mathbb{R}^{2\times L}.
	\label{eq:iq_input}
\end{equation}
For simplicity, a received signal segment, either in its complex sequence form or raw IQ matrix form, is denoted by $X$ in the following problem formulation.

\subsubsection{Cross-Domain AMC Task}
\label{sec:cross_domain_amc}

In this work, cross-domain AMC is formulated as modulation classification under domain-shifted channel conditions. 
The source and target domains share the same modulation label space but follow different signal distributions. 
The labeled source domain is denoted as 
$\mathcal{D}_s=\{(X_i^s,y_i^s)\}_{i=1}^{N_s}$, where $X_i^s$ is a received signal sample and $y_i^s\in\mathcal{Y}$ is its modulation label. 
The unlabeled target domain is denoted as 
$\mathcal{D}_t=\{X_j^t\}_{j=1}^{N_t}$, where $X_j^t$ is collected under different channel or hardware conditions. 
Thus, $P_s(X)\neq P_t(X)$ and $\mathcal{Y}_s=\mathcal{Y}_t=\mathcal{Y}$.

The objective is to learn a feature extractor $F(\cdot)$ and a classifier $C(\cdot)$ by exploiting $\mathcal{D}_s$ and $\mathcal{D}_t$, such that the expected target-domain risk is minimized:
\begin{equation}
	\min_{F,C}
	\;
	\epsilon_t(C\circ F)
	=
	\mathbb{E}_{(X,y)\sim P_t}
	\left[
	\ell
	\left(
	C(F(X)),y
	\right)
	\right],
	\label{eq:target_risk}
\end{equation}
where $\ell(\cdot,\cdot)$ denotes the classification loss. 
Since the target labels are unavailable during training, \eqref{eq:target_risk} cannot be directly optimized. 
This motivates knowledge-guided representation learning, where structural signal priors are introduced to learn features that are discriminative on the source domain and transferable to the target domain.

 \subsection{Signal Priors for Cross-Domain Modulation Representation}
 \label{sec:representation_analysis}
 
 Based on the signal model in \eqref{eq:signal_model}, the received signal involves multiple domain-dependent transformations. 
 Each signal representation can be viewed as a structural prior that emphasizes specific physical or statistical properties. 
 Analyzing these representations helps identify suitable priors for cross-domain representation learning.
 In this work, we examine five representations widely adopted in AMC: raw IQ samples, AP, DFT, ACF, and CD. 
 They describe the received signal from complementary perspectives, including waveform, amplitude--phase, spectral, correlation, and constellation-geometry domains.
 However, richer signal descriptions do not necessarily imply better cross-domain generalization. 
 A representation that retains detailed signal information may also preserve domain-specific distortions, whereas a more stable one may discard part of the modulation-discriminative information. 
 Therefore, candidate signal priors should be analyzed according to their discriminability, cross-domain stability, and complementarity.
 
Let $\psi_m(\cdot)$ denote the $m$-th signal representation constructed from a received signal segment $X$, where
$m \in \{\mathrm{IQ}, \mathrm{AP}, \mathrm{DFT}, \mathrm{ACF}, \mathrm{CD}\}$.
To organize the analysis, we characterize each representation from two aspects: modulation discriminability and domain discrepancy.

Modulation discriminability reflects whether different modulation formats can be well separated after the mapping $\psi_m(\cdot)$.
Conceptually, let $P_m^y := P(\psi_m(X)\mid y)$ denote the class-conditional distribution induced by $\psi_m(\cdot)$.
Then the class separability can be described as
\begin{equation}
	D_{\mathrm{cls}}(\psi_m)
	=
	\mathbb{E}_{y_i \ne y_j}
	\left[
	\operatorname{Div}
	\left(
	P_m^{y_i}, P_m^{y_j}
	\right)
	\right],
\end{equation}
where $\operatorname{Div}(\cdot,\cdot)$ denotes a generic distribution discrepancy measure.
A larger $D_{\mathrm{cls}}(\psi_m)$ indicates stronger separability among modulation classes.

Domain discrepancy characterizes how the source-target distribution mismatch is reflected in the representation space specified by $\psi_m(\cdot)$:
\begin{equation}
	D_{\mathrm{dom}}(\psi_m)
	=
	\operatorname{Div}
	\left(
	P_s(\psi_m(X)), P_t(\psi_m(X))
	\right).
\end{equation}
A smaller $D_{\mathrm{dom}}(\psi_m)$ indicates a smaller representation-level distribution shift between the source and target domains.
In the experiments, H-score and MMD are used as empirical indicators of class discriminability and domain discrepancy, respectively.
Together with the complementarity among different signal descriptions, they help support the choice of structural priors in the proposed framework.

Raw IQ samples provide the basic observation of the received complex baseband signal and preserve rich waveform-level modulation cues.
However, as shown in \eqref{eq:signal_model}, the received waveform is also affected by domain-dependent factors such as channel variations, noise, and transmission effects.
When only IQ samples are used, the model must directly separate modulation-intrinsic structures from domain-specific variations in the raw observation, which may increase its reliance on source-specific waveform patterns.
Therefore, although IQ is essential, additional structural priors are needed to provide complementary and more domain-stable inductive biases.

AP representation decomposes the received complex sequence into amplitude and phase trajectories:
\begin{equation}
	\psi_{\mathrm{AP}}(X_d)
	=
	\big[\, |X_d[n]|,\angle X_d[n] \,\big]_{n=1}^{L}.
	\label{eq:ap_representation}
\end{equation}
By reorganizing the received waveform into an explicit amplitude and phase view, AP makes amplitude and phase-dependent modulation structures more accessible.
The amplitude trajectory reflects envelope variations, while the phase trajectory captures temporal phase evolution.
These sequential amplitude-phase dynamics provide discriminative cues that are implicit in raw IQ samples but become more explicit in the AP representation.

While these amplitude-phase dynamics provide useful complementary cues, they are not sufficient to ensure domain stability by themselves.
For illustration, consider a simplified flat-fading case of \eqref{eq:signal_model}, where timing offset and noise are neglected and the channel is approximated by a complex gain $g_d$.
After sampling, the received sequence is expressed as

\begin{equation}
	X_d[n]
	\approx
	a_d g_d S[n]
	e^{j(\phi_d-2\pi \Delta f_d nT_s)},
	\label{eq:ap_simplified_signal}
\end{equation}
where $S[n]$ denotes the sampled transmitted sequence.
The phase component then satisfies
\begin{equation}
	\angle X_d[n]
	\approx
	\angle S[n]
	+
	\angle g_d
	+
	\phi_d
	-
	2\pi \Delta f_d nT_s .
	\label{eq:ap_phase_sensitivity}
\end{equation}
This relation shows that the observed phase trajectory contains both modulation-related phase evolution and domain-dependent phase terms, including channel phase, phase offset, and CFO-induced phase drift.
The amplitude trajectory can also be affected by amplitude scaling and channel gain.
Therefore, AP mainly serves as a complementary discriminative prior, and its limited domain stability motivates the use of additional domain-stable structural priors.

ACF captures the second-order temporal dependency of the received signal and provides a correlation-domain description.
For a time lag $\ell$, it is defined as
\begin{equation}
	\psi_{\mathrm{ACF}}(X_d)
	=
	R_{X_d}[\ell]
	=
	\mathbb{E}
	\left[
	X_d[n]X_d^*[n-\ell]
	\right].
	\label{eq:acf_representation}
\end{equation}

Compared with sample-level waveform representations, ACF describes the signal through second-order temporal statistics, providing a more regularized correlation-domain view of the received waveform.
The stability of ACF can be illustrated from several common domain-dependent transformations.

First, consider a constant phase rotation and amplitude scaling, i.e.,
$X_d[n]=a_d S[n]e^{j\phi_d}$, where $S[n]$ denotes the sampled transmitted sequence.
Then,
\begin{equation}
	\begin{aligned}
		R_{X_d}[\ell]
		&=
		\mathbb{E}
		\left[
		a_d S[n]e^{j\phi_d}
		a_d S^*[n-\ell]e^{-j\phi_d}
		\right] \\
		&=
		a_d^2 R_S[\ell].
	\end{aligned}
	\label{eq:acf_phase_invariance}
\end{equation}
Thus, the constant phase rotation is canceled by conjugate multiplication, and the effect of amplitude scaling can be further reduced by normalization:
\begin{equation}
	\widetilde{R}_{X_d}[\ell]
	=
	\frac{R_{X_d}[\ell]}{R_{X_d}[0]}.
	\label{eq:acf_normalization}
\end{equation}
For additive noise, let $X_d[n]=S[n]+N_d[n]$, where $S[n]$ and $N_d[n]$ are independent and $N_d[n]$ is zero-mean white noise.
Then,
\begin{equation}
	R_{X_d}[\ell]
	=
	R_S[\ell] + R_{N_d}[\ell],
	\label{eq:acf_noise}
\end{equation}
where $R_{N_d}[\ell]=0$ for $\ell\neq 0$.
Therefore, non-zero-lag ACF can suppress uncorrelated white noise while preserving the temporal dependency of the modulated signal.

ACF still has limitations under certain domain-dependent transformations.
For example, under the CFO term in \eqref{eq:signal_model}, if
$X_d[n]=S[n]e^{-j2\pi \Delta f_d nT_s}$, then
\begin{equation}
	\begin{aligned}
		R_{X_d}[\ell]
		&=
		\mathbb{E}
		\left[
		S[n]e^{-j2\pi \Delta f_d nT_s}
		S^*[n-\ell]e^{j2\pi \Delta f_d (n-\ell)T_s}
		\right] \\
		&=
		e^{-j2\pi \Delta f_d \ell T_s}R_S[\ell].
	\end{aligned}
	\label{eq:acf_cfo_effect}
\end{equation}
This shows that CFO is transformed into a lag-dependent phase factor in the correlation domain, rather than being completely removed.
Similarly, multipath fading can affect the second-order statistics of the received signal.
Under a quasi-static linear channel approximation, this effect can be expressed in the frequency domain as
\begin{equation}
	\Phi_{X^d}(f)=|H_d(f)|^2\Phi_S(f),
	\label{eq:acf_multipath_effect}
\end{equation}
where $\Phi_{X^d}(f)$ and $\Phi_S(f)$ denote the power spectral densities of $X_d[n]$ and $S[n]$, respectively.
These effects indicate that ACF is not fully invariant, but its domain variations are expressed in a structured correlation-domain form.
Therefore, ACF provides auxiliary correlation-domain information as a relatively domain-stable structural prior for cross-domain representation learning.

DFT provides a frequency-domain view of the received sequence and can offer complementary spectral cues for modulation classification:
\begin{equation}
	\psi_{\mathrm{DFT}}(X_d)
	=
	R_d[k]
	=
	\sum_{n=1}^{L}
	X_d[n]e^{-j2\pi kn/L}.
	\label{eq:dft_representation}
\end{equation}

However, spectral structures may also encode domain-dependent variations.
Under the CFO term in \eqref{eq:signal_model}, a simplified frequency-domain form can be written as
\begin{equation}
	R_d(f)
	\approx
	a_d e^{j\phi_d}
	H_d(f+\Delta f_d)
	S(f+\Delta f_d)
	+
	N_d(f),
	\label{eq:dft_domain_effect}
\end{equation}
where $S(f)$, $H_d(f)$, and $N_d(f)$ denote the spectra of the transmitted signal, channel response, and noise, respectively.
This relation shows that CFO shifts the spectral location, while the domain-dependent channel response reshapes the spectral envelope.
Sampling-rate mismatch and filtering may further alter the frequency-domain structure.
Thus, DFT is informative but may introduce domain-sensitive spectral bias when spectral distortions vary across domains.
Therefore, it is considered as a useful candidate representation, but is not selected as the final structural prior in this work.

CD representation describes the geometric distribution of complex samples on the IQ plane.
Let 
$\mathbf{z}_d[n]=\big(\mathrm{Re}\{X_d[n]\},\mathrm{Im}\{X_d[n]\}\big)$
denote the two-dimensional IQ point of the $n$-th sample.
When implemented as a density map, CD can be formulated as
\begin{equation}
	\psi_{\mathrm{CD}}^d(u,v)
	=
	\frac{1}{L}
	\sum_{n=1}^{L}
	1
	\left[
	\mathbf{z}_d[n]\in \mathcal{B}_{u,v}
	\right],
	\label{eq:cd_representation}
\end{equation}
where $\mathcal{B}_{u,v}$ denotes the $(u,v)$-th bin on the IQ plane.
CD can provide interpretable constellation-geometry patterns under well-synchronized and compensated conditions.
However, in cross-domain scenarios, phase offset, CFO, fading, timing mismatch, and noise may distort the constellation density distribution.
In addition, the density-map construction largely removes temporal ordering, which may weaken useful temporal evolution information.
Therefore, CD may introduce domain-sensitive geometric bias when synchronization and channel conditions vary across domains.

The above analysis shows that signal representations encode diverse prior information and exhibit different sensitivities to domain-dependent transformations.
Therefore, structural priors should be selected by jointly considering discriminability, cross-domain stability, and complementarity.
Although DFT and CD provide useful spectral and constellation-geometry cues, their structures are more closely coupled with domain-dependent impairments, making them informative candidates rather than final priors.
Accordingly, IQ, AP, and ACF are selected as a compact prior set for cross-domain representation learning.
Specifically, IQ preserves the fundamental waveform observation, AP makes amplitude--phase dynamics explicit, and ACF provides relatively stable second-order correlation statistics.
Their combination retains discriminative signal information while introducing complementary and domain-stable structural cues.
Based on this analysis, we design a dual knowledge and data-driven framework to learn, fuse, and align these prior-guided representations for cross-domain AMC.

\begin{figure*}
		\centering
		\includegraphics[width=1\linewidth]{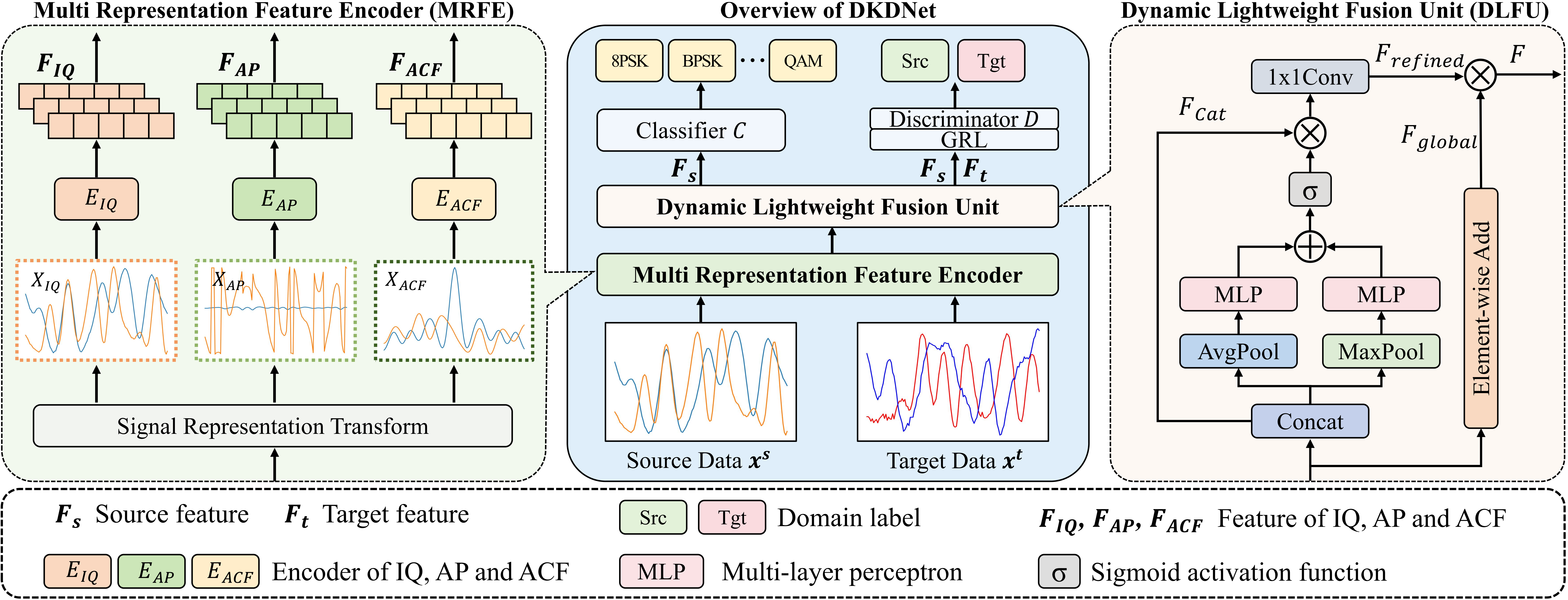}
		\caption{			
			Framework of the proposed DKDNet. 
			Source samples $x^s$ and target samples $x^t$ are first transformed into three representations, from which MRFE extracts representation-specific embeddings. Then, DLFU produces fused features $F_s$ and $F_t$. Subsequently, $F_s$ is fed into the classifier $C$ for modulation recognition, while both $F_s$ and $F_t$ are input to the discriminator $D$ for cross-domain alignment.		
			}
		\label{fig:pipeline}
\end{figure*}

\subsection{Overview of DKDNet}

DKDNet integrates structural signal priors with data-driven feature learning. Specifically, IQ, AP, and ACF are used as knowledge-guided inputs, which encode waveform structure, amplitude--phase evolution, and statistical periodicity, respectively.
MRFE processes each representation independently and maps them into a unified feature space, while DLFU adaptively fuses the extracted features. The fused features are jointly optimized by modulation classification and domain alignment objectives, enabling robust cross-domain generalization.
In this way, the signal priors provide structured and complementary inputs, and the network learns to extract, fuse, and align them for robust cross-domain modulation recognition. The overall architecture is shown in Fig.~\ref{fig:pipeline}, and the training procedure is presented in Algorithm~\ref{algo1}.

\subsection{Multi-Representation Feature Encoder}

Given the selected representations $X_{\mathrm{IQ}}$, $X_{\mathrm{AP}}$, and $X_{\mathrm{ACF}}$, the MRFE is designed to extract representation-specific features while mapping them into a unified latent space. After the representation construction described above, the received signal is represented as three real-valued matrices, i.e., $X_{\mathrm{IQ}}\in\mathbb{R}^{2\times L}$, $X_{\mathrm{AP}}\in\mathbb{R}^{2\times L}$, and $X_{\mathrm{ACF}}\in\mathbb{R}^{2\times L}$, where $L$ denotes the signal length. Although these inputs share the same matrix size, their channel semantics and statistical properties are different. Directly mixing them at the input level may therefore obscure representation-specific characteristics.

To address this issue, MRFE adopts three independent encoding branches to process the three representations separately. Specifically, input-level normalization is first applied to reduce scale differences among the inputs. The normalized representations are then fed into three parallel feature extractors, denoted as $E_{\mathrm{IQ}}$, $E_{\mathrm{AP}}$, and $E_{\mathrm{ACF}}$. These extractors share the same architecture but have independent parameters:
\begin{equation}
	F_m = E_m(X_m), \quad m \in \{\mathrm{IQ}, \mathrm{AP}, \mathrm{ACF}\},
\end{equation}
where $F_{\mathrm{IQ}}$, $F_{\mathrm{AP}}$, and $F_{\mathrm{ACF}}$ denote the extracted features. All three feature maps are projected into $\mathbb{R}^{B \times D}$, where $B$ is the batch size and $D$ is the feature dimension. Feature-level normalization is further applied to place them in a common embedding space, making the features comparable for subsequent fusion.

This branch-wise design enables each encoder to capture the characteristics of its corresponding representation while producing features with consistent dimensions. In our implementation, each feature extractor consists of four one-dimensional convolutional layers with kernel size $1\times3$, stride 1, and padding 1, producing feature maps with 16, 32, 64, and 2 channels, respectively. Each convolutional layer is followed by batch normalization and a LeakyReLU activation. A final LSTM layer with a hidden size of 128 is used to model temporal dependencies.
During domain adaptation, source and target samples share the same MRFE parameters, ensuring consistent feature extraction across domains.

\subsection{Dynamic Lightweight Fusion Unit}

After MRFE, the three feature vectors $F_{\mathrm{IQ}}$, $F_{\mathrm{AP}}$, and $F_{\mathrm{ACF}}$ lie in a unified feature space but may contribute differently to modulation classification and domain alignment. DLFU is designed to adaptively fuse these features with a lightweight dual-branch structure.

The first branch performs global additive fusion:
\begin{equation}
	F_{\mathrm{global}} = F_{\mathrm{IQ}} + F_{\mathrm{AP}} + F_{\mathrm{ACF}},
\end{equation}
where $F_{\mathrm{global}}\in\mathbb{R}^{B\times D}$. This branch preserves the overall information shared by the three representations.

The second branch estimates representation-wise feature importance. The three features are first stacked along the representation dimension:
\begin{equation}
	F_{\mathrm{cat}} =
	\mathrm{Concat}(F_{\mathrm{IQ}}, F_{\mathrm{AP}}, F_{\mathrm{ACF}}),
\end{equation}
where $F_{\mathrm{cat}}\in\mathbb{R}^{B\times 3\times D}$. 
Then, the pooling operations are performed along the feature dimension to obtain representation-level descriptors, which are then mapped and broadcast to generate $A \in \mathbb{R}^{B\times 3\times D}$.
\begin{align}
	A = \sigma \big(
	&\mathrm{MLP}(\mathrm{MaxPool}(F_{\mathrm{cat}})) \nonumber\\
	&+
	\mathrm{MLP}(\mathrm{AvgPool}(F_{\mathrm{cat}}))
	\big),
\end{align}
where $\sigma(\cdot)$ denotes the sigmoid activation function.

The importance score is applied to $F_{\mathrm{cat}}$, and a $1\times1$ convolution is used to reduce the representation dimension:
\begin{equation}
	F_{\mathrm{refined}}
	=
	\mathrm{Conv}_{1\times1}(F_{\mathrm{cat}}\odot A),
\end{equation}
where $F_{\mathrm{refined}}\in\mathbb{R}^{B\times D}$ and $\odot$ denotes element-wise multiplication.

Finally, the outputs of the two branches are combined to obtain the fused feature:
\begin{equation}
	F = F_{\mathrm{global}}\odot F_{\mathrm{refined}}.
\end{equation}
For source and target samples, the resulting fused features are denoted as $F_s$ and $F_t$, respectively.

This fusion strategy retains global complementary information while adaptively emphasizing informative components, providing a compact feature representation for subsequent classification and domain alignment.

\begin{algorithm}[t]
	\caption{Training process of DKDNet.}
	\label{algo1}
	\begin{algorithmic}[1]
		\STATE \textbf{Input:} Labeled source samples $\{(x_i^s,y_i^s)\}_{i=1}^{N_s}$, unlabeled target samples $\{x_j^t\}_{j=1}^{N_t}$, learning rate $lr$, and maximum epoch $K$.
		\STATE \textbf{Output:} Trained DKDNet.
		
		\STATE Initialize the parameters of MRFE, DLFU, classifier $C$, and discriminator $D$.
		
		\FOR{epoch $=1$ to $K$}
		\FOR{each source-target mini-batch}
		\STATE Construct $X_{\mathrm{IQ}}$, $X_{\mathrm{AP}}$, and $X_{\mathrm{ACF}}$ for source and target samples.
		\STATE Extract representation-specific features with MRFE.
		\STATE Fuse the extracted features with DLFU to obtain $F_s$ and $F_t$.
		\STATE Compute the source classification loss $\mathcal{L}_c$.
		\STATE Compute the domain-adversarial loss $\mathcal{L}_d$.
		\STATE Update all trainable parameters by optimizing $\mathcal{L}_{\mathrm{total}}$ with Adam.
		\ENDFOR
		\ENDFOR
		
		\STATE \textbf{Return:} Trained DKDNet.
	\end{algorithmic}
\end{algorithm}

\subsection{Modulation Classifier and Domain Discriminator}

DKDNet optimizes the fused representation using two objectives: source-domain modulation classification and source-target domain alignment. In our implementation, a modulation classifier $C$ is used for label prediction, and a domain discriminator $D$ with a Gradient Reversal Layer (GRL) is used for adversarial domain alignment.

\subsubsection{Modulation Classifier}

The modulation classifier predicts the modulation category from the fused source feature $F_s$. It consists of two fully connected layers, where the first layer is followed by dropout and a LeakyReLU activation. The output dimension is set to $C$, corresponding to the number of modulation classes. For source samples, the classification loss is defined as
\begin{equation}
	\mathcal{L}_{c}
	=
	-\frac{1}{N_s}
	\sum_{i=1}^{N_s}
	\sum_{c=1}^{C}
	1_{[y_i^s=c]}\log h_c(F_i^s),
\end{equation}
where $y_i^s$ is the label of the $i$-th source sample, $h_c(F_i^s)$ denotes the predicted probability of class $c$, and $1_{[y_i^s=c]}$ is the indicator function.

\subsubsection{Domain Discriminator}

The domain discriminator predicts whether a fused feature comes from the source or target domain. It has a similar structure to the modulation classifier but removes the dropout layer and outputs a binary domain prediction. A GRL is inserted before the discriminator. During backpropagation, GRL multiplies the gradients by $-\lambda$, encouraging the feature extractor to learn domain-invariant representations.
Let $F_i$ be the fused feature of the $i$-th sample and $d_i\in\{0,1\}$ its domain label, where 1 and 0 denote the source and target domains, respectively. With $N=N_s+N_t$, the domain-adversarial loss is
\begin{equation}
	\begin{aligned}
		\mathcal{L}_{d}
		=
		-\frac{1}{N}
		\sum_{i=1}^{N}
		\big[
		&d_i\log D(F_i)  \\
		&+(1-d_i)\log(1-D(F_i))
		\big].
	\end{aligned}
\end{equation}

The overall training objective is formulated as
\begin{equation}
	\mathcal{L}_{\mathrm{total}}
	=
	\mathcal{L}_{c}
	+
\mathcal{L}_{d},
\end{equation}
where the GRL reverses the gradient from $\mathcal{L}_{d}$ with a coefficient $\lambda$ when updating the feature extractor.

\begin{table}[b]
	\caption{Description of the RML2025 Series. MDS denotes the maximum Doppler shift in Hz.}
	\centering
	\renewcommand\arraystretch{1.5}
	\begin{tabular}{c|c|c|c|c}
		\hline
		\textbf{Dataset} & \makecell[c]{\textbf{MDS(Hz)}} & \textbf{Delay (ms)} & \textbf{Gains} &\textbf{ CFO / SRO}\\
		\hline
		AWGN &/ &/ &/  &/\\
		\hline
		Ri1 &4 &[0,0.9,1.7] &[1,0.8,0.3]   &/ \\
		\hline
		Ri2 &4 &[0,0.9,1.7] &[1,0.8,0.3] &50 (std 0.01)  \\		
		Ri3 &30 &\makecell[c]{[0,0.05,0.12,\\0.2,0.23,0.5,\\1.6,2.3,5]} &\makecell[c]{[0.8, 0.8,0.8,\\0.7,0.6,0.4,\\0.4,0.4,0.3]} &50 (std 0.01)  \\
		\hline
		Ray1 &1 &[0,0.9,1.7] &[1,0.8,0.3]  &/  \\
		
		Ray2 &3 &[0,0.9,1.7] &[1,0.8,0.3] &50 (std 0.01)  \\	
		Ray3 &30 &\makecell[c]{[0,0.05,0.12,\\0.2,0.23,0.5,\\1.6,2.3,5]} &\makecell[c]{[0.8, 0.8,0.8,\\0.7,0.6,0.4,\\0.4,0.4,0.3]}  &50 (std 0.01) \\
		\hline
	\end{tabular}
	\label{table:dataset}
\end{table}

\section{EXPERIMENTS}

\subsection{Experimental Setup}
We evaluate DKDNet on two datasets with complementary purposes.
The proposed RML2025 Series is used as the main benchmark, as it provides controlled cross-domain AMC scenarios with progressively intensified channel impairments, including multipath fading, CFO, SRO, and Doppler shift.
It is used for prior selection, method comparison, ablation studies, backbone compatibility, UDA compatibility, and sample-efficiency analysis.
In addition, the public RML22 benchmark\cite{RML22} is used as an external dataset to further examine the generality of DKDNet beyond the proposed simulated benchmark.
For both datasets, we follow the UDA setting, where labeled source-domain samples and unlabeled target-domain samples are available during training, while target-domain labels are used only for evaluation.

Experiments are implemented in PyTorch 2.0.1 and Python 3.8 on a single NVIDIA GeForce RTX 3090 GPU with 24 GB memory.
The model is trained for 200 epochs using Adam with a fixed learning rate of 5e-4 and a batch size of 512 for both source and target domains.
Each iteration uses synchronized sampling, processing one mini-batch from each domain.
Convolutional and linear layers are initialized by Kaiming and Glorot initialization, respectively.
For regularization, 1D batch normalization follows each convolutional layer, and dropout with a rate of 0.5 is applied after each linear layer.
All experiments use five random seeds, and results are reported as mean and standard deviation.
For all reported improvements over the corresponding baselines, statistical significance is evaluated using two-sided paired $t$-tests across the five random seeds. The improvements marked or discussed as performance gains are statistically significant with $p<0.05$.

\subsection{Construction of the RML2025 Series Datasets}
Existing AMC datasets mainly support in-domain evaluation or contain limited cross-domain variations. 
Commonly used RadioML datasets~\cite{o2016radio} and private simulated datasets are therefore insufficient for systematically evaluating cross-domain AMC methods under controlled and progressively intensified channel impairments.
To address this limitation, we construct the RML2025 Series, a set of simulated domains for cross-domain AMC evaluation with structured source-target domain shifts.
The considered impairments include noise, multipath fading, carrier frequency offset (CFO), sampling rate offset (SRO), and Doppler shift, which affect the temporal, spectral, and constellation structures of received signals.

The RML2025 Series is built upon the widely adopted RadioML codebase\footnote{\url{https://github.com/radioML/dataset}} and comprises seven subsets, each corresponding to a distinct channel condition. The sampling rate is set to 200 kHz, with 8 samples per symbol, corresponding to a symbol rate of 25 kSymbols/s. Each dataset contains 11 modulation types, namely 8PSK, AM-DSB, AM-SSB, BPSK, CPFSK, GFSK, PAM4, QAM16, QAM64, QPSK, and WB-FM. The SNR ranges from -20 dB to 18 dB with a step size of 2 dB. For each dataset, each modulation contains 1000 samples at each SNR level, resulting in 220k samples. Each sample consists of 128 complex-valued IQ points. Energy normalization is applied to remove trivial energy-scale differences.

The seven subsets include AWGN, Ri1--Ri3, and Ray1--Ray3.
AWGN contains only additive white Gaussian noise and is used as the basic source domain in most experiments.
For Rician and Rayleigh channels, three impairment levels with increasing severity are considered.
Ri1/Ray1 mainly include multipath fading and Doppler effects, Ri2/Ray2 further introduce CFO and SRO, and Ri3/Ray3 adopt more complex multipath profiles and larger Doppler shifts.
This design provides progressively more challenging target domains for evaluating cross-domain generalization.
The detailed configurations are summarized in Table~\ref{table:dataset}, and the datasets are publicly released on GitHub\footnote{\url{https://github.com/FireTracer/RML2025-Series}}.

The impairments are configured as follows.
The AWGN subset contains only additive white Gaussian noise without multipath effects.
The Rician subsets model propagation with a dominant line-of-sight path and scattered multipath components, with the K-factor fixed to 4, while the Rayleigh subsets represent non-line-of-sight scattered propagation without a dominant path.
Different delay spreads and path-gain profiles are adopted for the Rician and Rayleigh subsets to increase domain diversity.
CFO is introduced in Ri2, Ri3, Ray2, and Ray3 to model oscillator mismatch, with a nominal offset of 50 Hz and a random perturbation of standard deviation 0.01 Hz.
Given the symbol rate of 25 kSymbols/s, the normalized CFO is $50/25000=0.002$.
SRO is applied to the same subsets to model sampling-clock mismatch and is set to 50 Hz relative to the 200 kHz sampling rate, corresponding to 250 ppm.
Doppler shift models mobility-induced frequency variation and is specified by the maximum Doppler shift in the fading channel model.

\begin{table}[t]
	\centering
	\caption{
		Empirical comparison of candidate single- and multi-representation signal priors under the AWGN$\rightarrow$Ri1 adaptation setting.
		H-score, MMD, adapted target-domain accuracy, and UDA gain are reported as mean$\pm$std over five random seeds. All reps. denotes the combination of IQ, AP, DFT, ACF, and CD.
	}
	\label{tab:rep_comparison}
	\setlength{\tabcolsep}{3.5pt}
	\renewcommand{\arraystretch}{1.05}
	\resizebox{\columnwidth}{!}{
		\begin{tabular}{lcccc}
			\toprule
			Representation 
			& H-score $\uparrow$
			& MMD$_{\times 10^{-3}}$ $\downarrow$
			& Acc. (\%) $\uparrow$
			& Gain (\%) $\uparrow$
			\\
			\midrule
			IQ  
			& \underline{0.406$\pm$0.019} 
			& 4.9$\pm$0.5     
			& \textbf{42.84$\pm$0.87}  
			& \underline{7.25$\pm$0.50}     
			\\
			AP  
			& 0.321$\pm$0.028    
			& 5.7$\pm$0.3  
			& \underline{40.80$\pm$1.04}   
			& \textbf{12.19$\pm$1.04}   
			\\
			DFT 
			& 0.299$\pm$0.026   
			& 3.2$\pm$0.8   
			& 36.26$\pm$0.82     
			& 4.75$\pm$0.44       
			\\
			ACF 
			& \textbf{0.454$\pm$0.043}  
			& \textbf{0.9$\pm$0.4}     
			& 38.48$\pm$0.09    
			& 0.78$\pm$0.18    
			\\
			CD 
			& 0.062$\pm$0.003   
			& \underline{1.8$\pm$0.1}  
			& 26.04$\pm$0.08 
			& 2.97$\pm$0.22     
			\\
			\midrule
			IQ+AP 
			& 0.353$\pm$0.016   
			& 7.2$\pm$1.5   
			& 43.82$\pm$2.69 
			& 9.70$\pm$3.64     
			\\
			IQ+DFT 
			& 0.356$\pm$0.030   
			& 5.3$\pm$0.9   
			& 44.11$\pm$0.45  
			& 10.64$\pm$0.95     
			\\
			IQ+ACF 
			& 0.346$\pm$0.013   
			& \textbf{3.0$\pm$0.2}   
			& 44.72$\pm$0.29  
			& 8.58$\pm$0.69     
			\\
			AP+ACF 
			& 0.292$\pm$0.012   
			& 4.8$\pm$0.7   
			& 40.26$\pm$0.92
			& 8.68$\pm$1.43     
			\\
			IQ+AP+ACF 
			& 0.376$\pm$0.015   
			& \underline{3.5$\pm$0.3}   
			& \textbf{47.58$\pm$0.92}  
			& \textbf{12.83$\pm$1.40}     
			\\
			IQ+DFT+ACF 
			& \underline{0.378$\pm$0.037}   
			& 5.3$\pm$0.7  
			& 43.77$\pm$3.16 
			& 9.18$\pm$2.34     
			\\
			IQ+AP+DFT 
			& 0.343$\pm$0.025   
			& 5.4$\pm$0.9   
			& \underline{46.99$\pm$1.89} 
			& \underline{12.40$\pm$1.78}     
			\\
			IQ+AP+CD 
			& 0.377$\pm$0.039   
			& 6.3$\pm$1.3  
			& 43.67$\pm$1.51  
			& 9.28$\pm$0.95     
			\\
			All reps.
			& \textbf{0.437$\pm$0.022}   
			& 5.4$\pm$0.4   
			& 45.61$\pm$1.23 
			& 9.03$\pm$0.94     
			\\
			\bottomrule
		\end{tabular}
	}
\end{table}

\subsection{Empirical Justification of Prior Selection}
\label{sec:input_analyse}

This subsection provides an empirical justification for the input selection of DKDNet based on the representation analysis in Section~III-B.
The evaluation is conducted under the AWGN$\rightarrow$Ri1 adaptation setting, with AWGN and Ri1 used as the source and target domains, respectively.
H-score~\cite{muandet2013domain} and MMD~\cite{gretton2012kernel} are used to measure modulation discriminability and source-target discrepancy, while the UDA gain is computed over the corresponding source-only target accuracy.
For implementation, one-dimensional representations are processed by one branch of MRFE, and CD is processed by a lightweight 2D CNN followed by a linear projection layer.
Multi-representation inputs are fused by linear addition to isolate the effect of prior selection from the proposed DLFU.
All reps. denotes the combination of IQ, AP, DFT, ACF, and CD.
The results are summarized in Table~\ref{tab:rep_comparison}.

The single-representation results show clear differences among the five priors. 
IQ achieves the best adapted accuracy among single representations, reaching $42.84\%$, indicating that raw waveform samples retain rich information for modulation classification.
However, its MMD is relatively large, indicating that IQ features are still sensitive to channel variations. AP obtains the largest UDA gain among single representations, i.e., $12.19\%$, suggesting that its amplitude--phase structure provides useful cues that can be effectively realigned by UDA. ACF achieves the highest H-score and the smallest MMD among single representations, with an MMD of $0.9\times10^{-3}$, showing that its correlation-domain structure is both well separated and relatively stable across domains.
Nevertheless, its adapted accuracy is only $38.48\%$, indicating that such stability is insufficient without detailed waveform-level information.
DFT and CD achieve lower adapted accuracies, suggesting that spectral shifts and constellation-geometry distortions may reduce their effectiveness under this cross-domain setting.

The multi-representation results further show that complementary priors are beneficial, but simply adding more representations is not always effective. IQ+AP+ACF achieves the best adapted accuracy of $47.58\%$ and the largest UDA gain of $12.83\%$, outperforming all single-representation inputs and other tested combinations. This supports the use of IQ, AP, and ACF as the prior-guided inputs of DKDNet: IQ provides detailed waveform information, AP introduces explicit amplitude--phase dynamics, and ACF contributes relatively stable correlation-domain statistics. In contrast, adding DFT or CD does not consistently improve performance. 
Although the combination of all representations obtains the highest H-score of $0.437$, its adapted accuracy drops to $45.61\%$, lower than IQ+AP+ACF. 
This indicates that stronger class separability alone does not guarantee better cross-domain adaptation. Domain-sensitive or redundant priors may instead hinder feature alignment. Therefore, IQ, AP, and ACF are selected as a compact and complementary prior set for DKDNet.

\begin{table}[t]		
	\centering
	\caption{
		Average target-domain accuracy of UDA experiments under different source-target configurations.
		Results are reported as mean$\pm$std over five random seeds.
	}
	\label{table:DA_alldataset}
	\renewcommand{\arraystretch}{1.25}
	\setlength{\tabcolsep}{4pt}
	\resizebox{\columnwidth}{!}{
		\begin{tabular}{l|l|cc|cc}
			\toprule
			\multirow{2}{*}{$\boldsymbol{D_s}$} 
			& \multirow{2}{*}{$\boldsymbol{D_t}$} 
			& \multicolumn{2}{c}{Acc. (\%) $\uparrow$ (-20 dB$\sim$18 dB)}
			& \multicolumn{2}{c}{Acc. (\%) $\uparrow$ (0 dB$\sim$18 dB)} \\ 
			\cmidrule(lr){3-4} \cmidrule(lr){5-6}
			& & Source Only & DKDNet & Source Only & DKDNet \\
			\midrule
			\multirow{6}{*}{AWGN}  
			& Ri1  & 33.89$\pm$0.98 & 48.93$\pm$0.54 & 47.40$\pm$1.83 & 76.67$\pm$1.25 \\
			& Ri2  & 35.65$\pm$0.55 & 44.76$\pm$2.17 & 51.75$\pm$0.92 & 59.46$\pm$4.15 \\
			& Ri3  & 35.86$\pm$0.52 & 44.91$\pm$2.09 & 49.37$\pm$0.95 & 66.63$\pm$4.03 \\
			\cmidrule(lr){2-6}
			& Ray1 & 40.39$\pm$0.79 & 49.74$\pm$2.26 & 56.94$\pm$1.64 & 75.97$\pm$4.26 \\
			& Ray2 & 37.42$\pm$0.57 & 46.76$\pm$3.42 & 52.85$\pm$1.04 & 71.32$\pm$6.71 \\
			& Ray3 & 35.37$\pm$0.69 & 43.40$\pm$2.65 & 48.77$\pm$0.99 & 63.97$\pm$5.34 \\
			\midrule
			Ri1  & Ray3 & 49.00$\pm$0.32 & 52.28$\pm$0.73 & 70.29$\pm$0.42 & 78.72$\pm$1.21 \\
			Ray1 & Ri3  & 49.81$\pm$0.37 & 52.10$\pm$0.39 & 71.37$\pm$0.49 & 78.20$\pm$0.64 \\
			\bottomrule
		\end{tabular}
	}
\end{table}

\begin{figure}
	\centering	
	\includegraphics[width=1\linewidth]{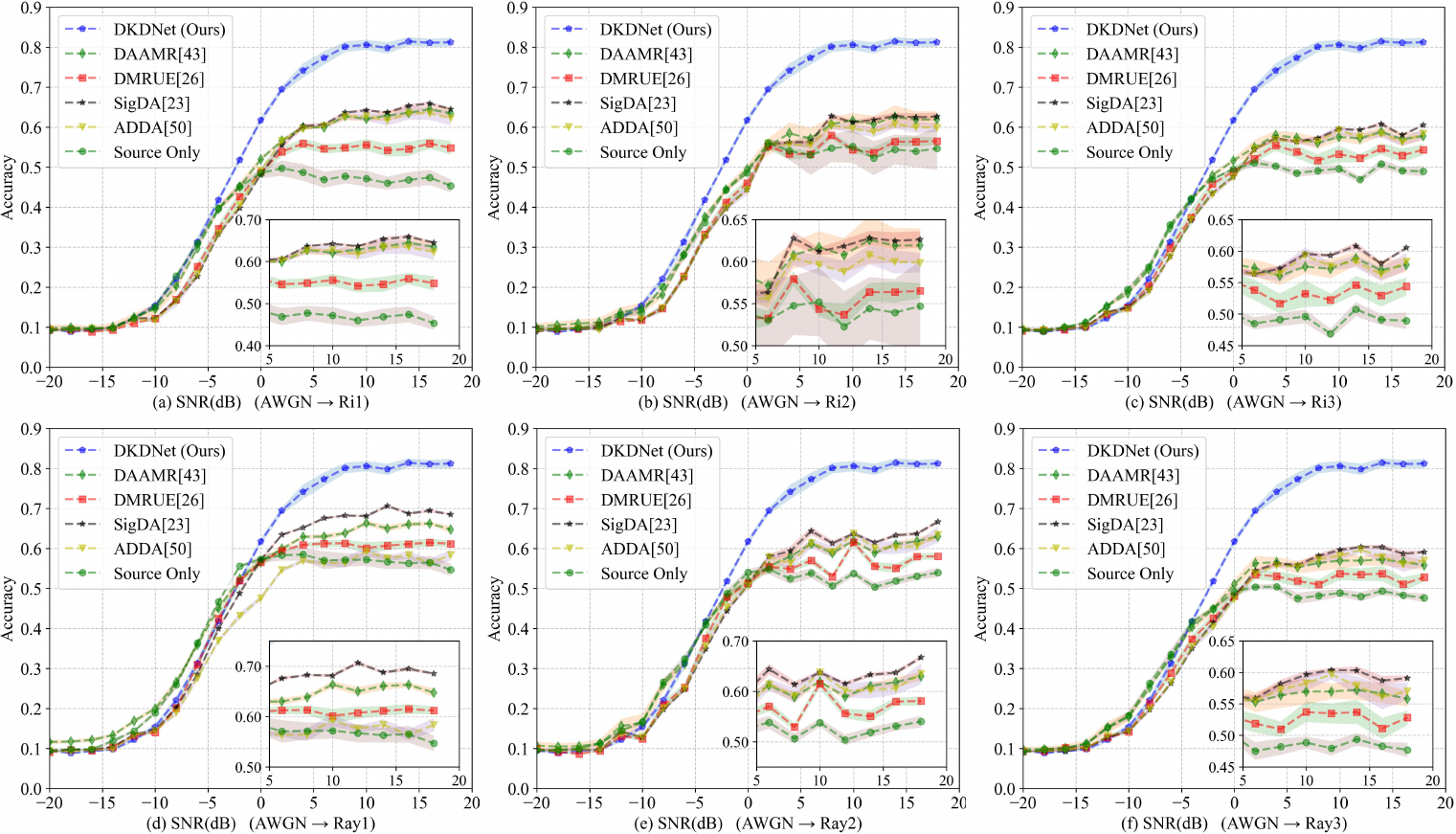}
	\caption{		
			Performance comparison between DKDNet and baseline methods across all SNR levels.
			AWGN is used as the source domain, and the target domains are: (a) Ri1, (b) Ri2,
			(c) Ri3, (d) Ray1, (e) Ray2, and (f) Ray3.
		}
	\label{fig:compare_dataset}
\end{figure}

\subsection{Effectiveness of the Proposed DKDNet Framework}

Table~\ref{table:DA_alldataset} reports the average target-domain accuracy under different source-target configurations. 
Source Only denotes the model trained only on the labeled source domain and directly tested on the target domain, while DKDNet denotes the proposed UDA framework.
For clarity, results are reported over all SNR levels and over the high-SNR range.

DKDNet consistently improves over Source Only in all evaluated settings.
When AWGN is used as the source domain, the target domains contain Rician or Rayleigh fading with different impairment levels, leading to relatively large domain shifts.
In these cases, DKDNet improves the all-SNR accuracy by 8.03--15.04 percentage points and the high-SNR accuracy by 7.71--29.27 percentage points.
The improvement is more pronounced in the high-SNR range, where modulation-related structures are less obscured by noise and can be more effectively aligned.
When the source domain is also impaired, i.e., Ri1$\rightarrow$Ray3 and Ray1$\rightarrow$Ri3, Source Only already achieves higher target accuracy, leaving less room for adaptation.
Even in these milder shifts, DKDNet still improves the all-SNR accuracy by 3.28 and 2.29 percentage points, respectively, and the high-SNR accuracy by 8.43 and 6.83 percentage points.
These results show that DKDNet is effective not only for AWGN-source adaptation, but also when the source domain contains fading impairments.

Fig.~\ref{fig:compare_dataset} further compares DKDNet with representative UDA methods under six AWGN-source adaptation tasks.
The compared methods include DAAMR~\cite{li2021domain}, DMRUE~\cite{li2019deep}, and SigDA~\cite{SigDA}, which are designed for AMC, as well as ADDA~\cite{ADDA}, a classical UDA method.
For fair comparison, the network structure for a single IQ input is kept consistent with one branch of MRFE.
When the SNR is above approximately $-4$ dB, DKDNet generally achieves the best performance across the six target domains.
For example, in the AWGN$\rightarrow$Ri1 task, DKDNet exceeds the strongest baseline by about 14 percentage points at 18 dB.
At very low SNR levels, all methods show limited performance because severe noise corruption obscures modulation-discriminative structures.
These results indicate that the performance gain of DKDNet comes not only from adversarial domain alignment, but also from the prior-guided multi-representation learning introduced before alignment.

\begin{table}[t]
	\centering
	\caption{
		Target-domain accuracy on the public RML22 benchmark under the RML22-AWGN$\rightarrow$RML22-full adaptation setting.
		Results are reported as mean$\pm$std over five random seeds.
	}
	\label{tab:rml22}
	\renewcommand{\arraystretch}{1.25}
	\setlength{\tabcolsep}{4pt}
	\resizebox{\columnwidth}{!}{
		\begin{tabular}{l|c|c}
			\toprule
			Method 
			& Acc. (\%) $\uparrow$ (-20 dB$\sim$18 dB)
			& Acc. (\%) $\uparrow$ (0 dB$\sim$18 dB) \\
			\midrule
			Source Only 
			& 23.91$\pm$0.41 
			& 28.79$\pm$0.66 \\
			\midrule
			DAAMR~\cite{li2021domain} 
			& 38.07$\pm$1.75 
			& 50.42$\pm$3.26 \\
			DMRUE~\cite{li2019deep} 
			& 31.76$\pm$0.77 
			& 42.69$\pm$1.04 \\
			SigDA~\cite{SigDA} 
			& 32.97$\pm$2.19 
			& 44.47$\pm$3.35 \\
			ADDA~\cite{ADDA} 
			& 28.62$\pm$0.86 
			& 37.87$\pm$1.80 \\
			\midrule
			DKDNet 
			& \textbf{42.05$\pm$3.94} 
			& \textbf{56.52$\pm$5.71} \\
			\bottomrule
		\end{tabular}
	}
\end{table}

In addition to the RML2025 Series, we further evaluate DKDNet on the public RML22 benchmark to examine its generality under another cross-domain AMC setting. Specifically, the RML22-AWGN subset is used as the source domain, while the full RML22 dataset, which contains mixed signal impairments, is used as the target domain. The results are summarized in Table~\ref{tab:rml22}.
DKDNet achieves the best performance on this public benchmark, reaching $42.05\%$ over all SNR levels and $56.52\%$ under high-SNR conditions. Compared with Source Only, DKDNet improves the target-domain accuracy by $18.14$ and $27.73$ percentage points, respectively. This indicates that the proposed prior-guided adaptation framework remains effective when the target domain contains mixed channel impairments. Compared with the strongest baseline DAAMR, DKDNet further improves the all-SNR accuracy from $38.07\%$ to $42.05\%$ and the high-SNR accuracy from $50.42\%$ to $56.52\%$. These results further demonstrate the generality and robustness of DKDNet under different cross-domain AMC settings.

\begin{figure}
	\centering
	\includegraphics[width=1\linewidth]{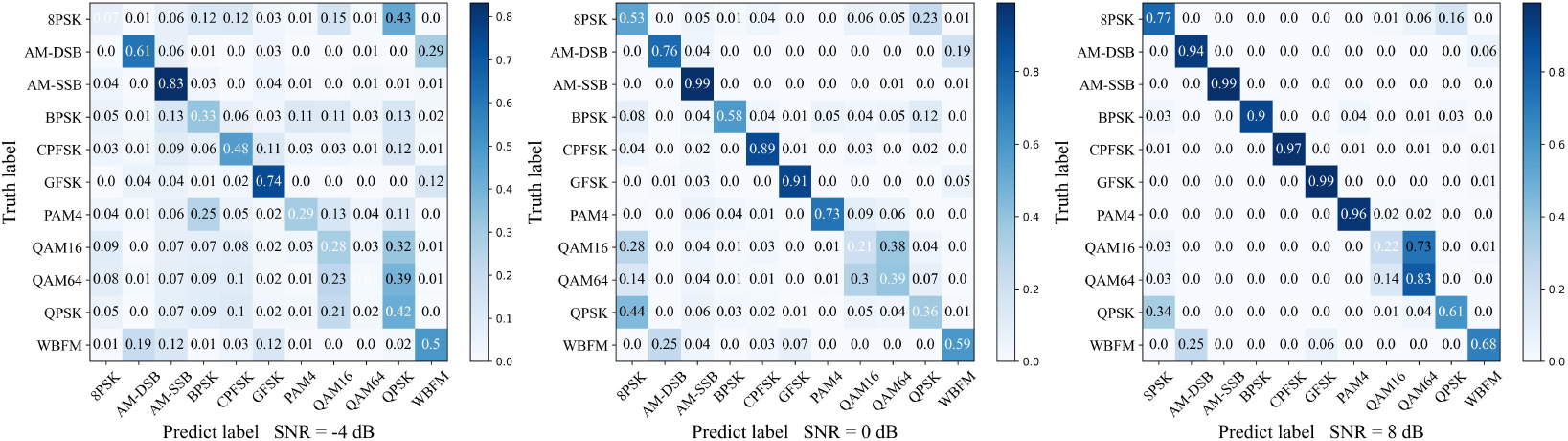}
	\caption{
		Confusion matrices of DKDNet under the AWGN$\rightarrow$Ri1 adaptation setting at $-4$, $0$, and $8$ dB.
	}
	\label{fig:confusion}
\end{figure}

\begin{figure}[h]		
		\centering
		\includegraphics[width=0.95\linewidth]{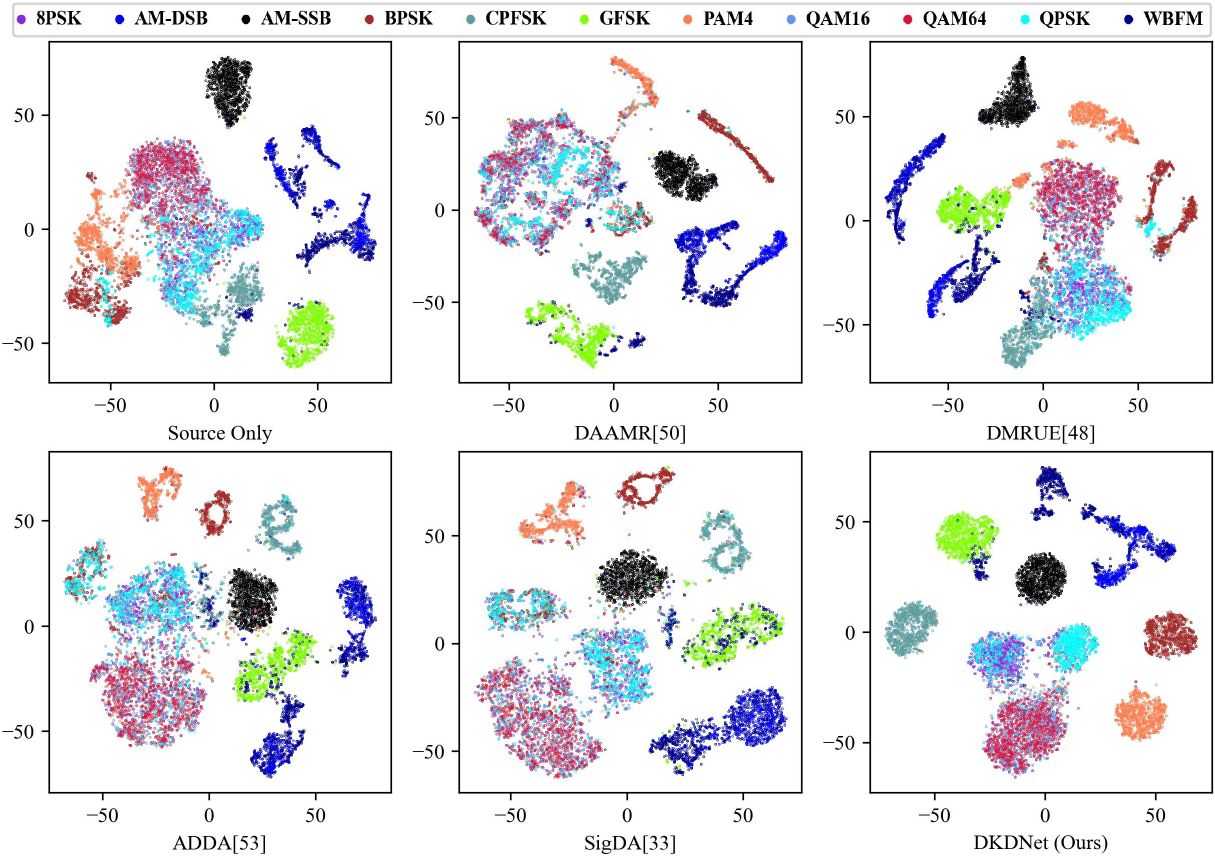}
		
		\caption{
			Class-level t-SNE visualization of target-domain features under the AWGN$\rightarrow$Ri1 adaptation setting.
		}		
		\label{fig:tsne_6}
\end{figure}

\begin{figure}[h]	
		\centering
		\includegraphics[width=0.8\linewidth]{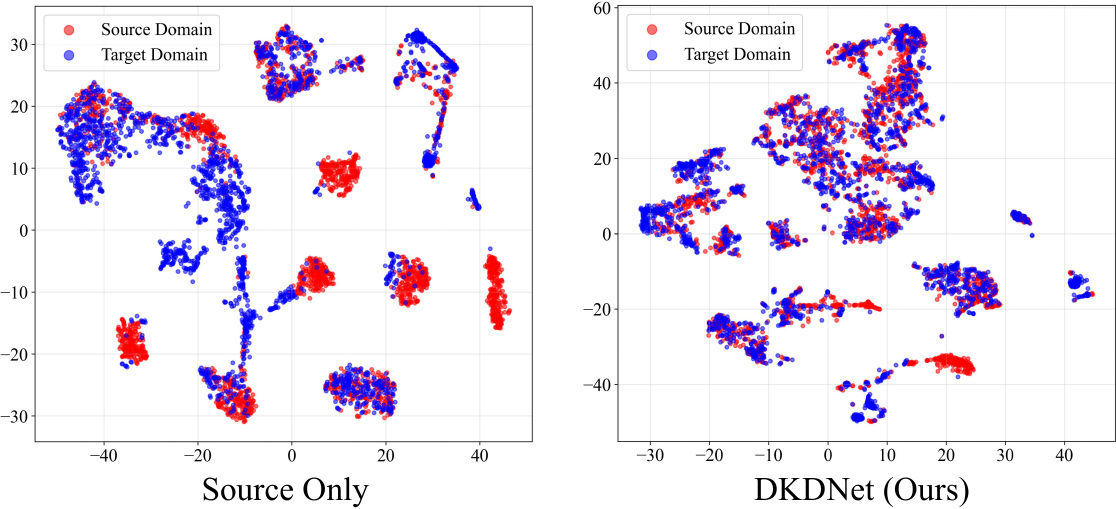}
		\caption{
			Domain-level t-SNE visualization before and after adaptation under the AWGN$\rightarrow$Ri1 setting.
		}
		\label{fig:tsne2}
\end{figure}

\subsection{Feature-Level and Class-Level Analysis}

We visualize class-level predictions and feature distributions under the AWGN$\rightarrow$Ri1 adaptation setting.
For each SNR level, the target-domain test set contains 200 samples per class, i.e., 2200 samples in total.
The confusion matrices are computed at $-4$, $0$, and $8$ dB and are row-normalized by the number of true samples in each class.
For t-SNE visualization, 2200 target-domain samples at 18 dB are used for class-level analysis, and 2200 source-domain samples together with 2200 target-domain samples at 18 dB are used for domain-level analysis.

Fig.~\ref{fig:confusion} shows that the diagonal entries become more dominant as the SNR increases, indicating improved target-domain class separability when modulation-related structures are less corrupted by noise.
At low SNRs, several modulation types are still confused due to severe noise corruption.
At higher SNRs, most classes are well recognized, and the remaining errors mainly occur between modulation formats with similar signal structures, such as QAM16/QAM64, 8PSK/QPSK, and AM-DSB/WBFM.
This trend is consistent with the SNR-wise results in Fig.~\ref{fig:compare_dataset}.
Fig.~\ref{fig:tsne_6} further compares the target-domain feature distributions learned by Source Only, representative UDA methods, and DKDNet.
Compared with the baselines, DKDNet produces more compact intra-class clusters and clearer inter-class separation, suggesting that the introduced signal priors help preserve modulation-discriminative structures in the target domain.
In addition, Fig.~\ref{fig:tsne2} shows stronger overlap between source and target features after adaptation, indicating reduced source-target discrepancy.
These visualizations demonstrate that DKDNet improves target-domain performance by learning features that are both discriminative and better aligned across domains.

\subsection{Ablation and Controlled Studies}

We conduct ablation and controlled studies under the AWGN$\rightarrow$Ri1 adaptation setting to analyze the source of DKDNet's performance gains.
The experiments focus on three aspects: the role of domain-stable priors, the effectiveness of the fusion strategy, and the complementarity of the selected representations.
All results are reported as mean$\pm$std over five random seeds.

\begin{table}[t]		
	\centering
	\caption{
		Controlled study on the source of DKDNet's performance gain under the AWGN$\rightarrow$Ri1 adaptation setting.
		P1 and P2 are linear projections of IQ with the same dimensionality as AP and ACF, without introducing additional signal-domain priors.
		Results are reported as mean$\pm$std over five random seeds.
	}
	\label{table:params}
	\renewcommand{\arraystretch}{1.25}
	\setlength{\tabcolsep}{3.5pt}
	\resizebox{\columnwidth}{!}{
		\begin{tabular}{l|cccc}
			\toprule
			Method & FLOPs & Params & Acc. (\%) $\uparrow$ & MMD $\downarrow$ \\
			\midrule
			DKDNet 
			& 29.66M 
			& 237.24K 
			& \textbf{48.93$\pm$0.54}
			& \textbf{0.0005$\pm$0.0003} \\
			\midrule
			IQ-tiny 
			& 9.85M 
			& 85.06K 
			& 42.84$\pm$0.87
			& 0.0012$\pm$0.0003 \\
			
			IQ-medium 
			& 98.21M 
			& 749.61K 
			& 45.24$\pm$0.41
			& 0.0021$\pm$0.0005 \\
			
			IQ-large 
			& 556.34M 
			& 4.35M 		
			& 45.72$\pm$0.58
			& 0.0014$\pm$0.0001 \\
			\midrule	
			Concatenation 
			& 29.66M 
			& 237.24K$^{\ast}$
			& \underline{47.45$\pm$1.62}
			& \underline{0.0007$\pm$0.0004} \\
			
			Linear Fusion 
			& 29.66M 
			& 237.24K$^{\ast}$
			& 46.79$\pm$0.97
			& 0.0012$\pm$0.0006 \\
			\midrule
			IQ+DFT+CD 
			& 29.66M 
			& 237.24K
			& 41.42$\pm$1.02
			& 0.0201$\pm$0.0053 \\
			
			IQ+P1+P2 
			& 29.66M 
			& 237.24K
			& 33.04$\pm$2.05
			& 0.0209$\pm$0.0056 \\
			\bottomrule
		\end{tabular}
	}
\end{table}

\subsubsection{Controlled Study on the Source of Performance Gain}

We first conduct controlled experiments to examine three possible factors: model capacity, fusion strategy, and prior quality.
First, IQ-tiny, IQ-medium, and IQ-large use only IQ inputs with different encoder scales, which evaluates whether increasing the capacity of an IQ-only model can reproduce the gain of DKDNet. 
Second, Concatenation and Linear Fusion replace DLFU with simpler fusion operations while keeping the selected IQ, AP, and ACF inputs, which assesses the effect of adaptive fusion. 
Third, AP and ACF are replaced by DFT and CD, or by two projected IQ views denoted as P1 and P2, to examine whether the gain comes from meaningful signal priors rather than merely increasing the number of input branches. 
IQ+P1+P2 is an information-matched baseline that uses two IQ projections with the same dimensionality as AP and ACF, matching DKDNet's input size without adding explicit signal-domain priors.
The results are summarized in Table~\ref{table:params}. The asterisk in Table~\ref{table:params} indicates that the parameter difference from DKDNet is negligible.

As shown in Table~\ref{table:params}, DKDNet achieves the highest target-domain accuracy of $48.93\%$ and the lowest MMD of $0.0005$, indicating stronger target-domain recognition performance and a smaller source-target feature discrepancy. 
The IQ-only baselines show that model capacity alone cannot explain this improvement: even IQ-large, with $556.34$M FLOPs and $4.35$M parameters, achieves only $45.72\%$ accuracy, which is still lower than DKDNet. 

The fusion variants further show that effective fusion contributes to the final performance. 
Concatenation and Linear Fusion achieve $47.45\%$ and $46.79\%$, respectively, which are lower than DKDNet but still competitive. 
This indicates that the selected representations provide useful complementary information, while DLFU further improves their integration through adaptive feature weighting. 

The prior-quality controls provide more direct evidence for the role of domain-stable priors. 
When AP and ACF are replaced by DFT and CD, the accuracy drops to $41.42\%$ and the MMD increases to $0.0201$. 
The information-matched IQ+P1+P2 baseline further decreases the accuracy to $33.04\%$ with an MMD of $0.0209$, despite matching DKDNet's input size and branch structure. 
These results show that simply increasing input dimensionality or adding IQ-derived branches does not improve cross-domain generalization. 
Overall, DKDNet's improvement mainly comes from incorporating stable and complementary signal priors, while the proposed fusion strategy further enhances their effective integration.

\begin{table}[t]
	\centering
	\renewcommand\arraystretch{1.15}
	\setlength{\tabcolsep}{2.5pt}
		\caption{
			Ablation study of fusion strategies under the AWGN$\rightarrow$Ri1 adaptation setting.
			All variants use the same IQ, AP, and ACF inputs.
			Results are reported as mean$\pm$std over five random seeds.
		}
		\begin{tabular}{l|ccccc}
			\toprule
			{Fusion} &
			{DLFU} &
			{Plus} &
			{Cat} &
			{Attention} &
			{Early} \\
			\midrule
			Acc. (\%) &
			\textbf{48.93$\pm$0.54} &
			46.79$\pm$0.97 &
			47.45$\pm$1.62 &
			47.61$\pm$0.95 &
			40.66$\pm$2.67 \\
			\bottomrule
		\end{tabular}
		\label{table:fusion_ablation}	
\end{table}

\subsubsection{Effect of Fusion Strategy}

To evaluate DLFU, we compare it with several simplified fusion strategies under the AWGN$\rightarrow$Ri1 adaptation setting. 
All variants use the same IQ, AP, and ACF inputs and representation-specific encoders, differing only in the fusion strategy. 
The compared variants include element-wise addition without adaptive weighting (Plus), feature concatenation with linear projection (Cat), attention-only fusion (Attention), and input-level concatenation before feature extraction (Early).
As shown in Table~\ref{table:fusion_ablation}, Early fusion performs the worst, achieving only $40.66\%$ accuracy. 
This suggests that directly mixing heterogeneous representations may obscure representation-specific structures. 
Feature-level fusion is more effective, with Plus, Cat, and Attention reaching $46.79\%$, $47.45\%$, and $47.61\%$, respectively. 
DLFU achieves the best accuracy of $48.93\%$, outperforming the attention-only variant by $1.32$ percentage points. 
These results show that combining global additive fusion with adaptive feature refinement enables more effective integration of complementary signal representations.

\begin{table}[t]
		\caption{
			Ablation study of signal representation combinations under the AWGN$\rightarrow$Ri1 adaptation setting.
			Results are reported as mean$\pm$std over five random seeds across all SNR levels.
		}
		\centering
		\renewcommand\arraystretch{1.15}

		\begin{tabular}{
				p{1.2cm}<{\centering}
				p{1.2cm}<{\centering}
				p{1.2cm}<{\centering}|
				p{2cm}<{\centering}
			}
			\toprule
			{IQ} & {AP} & {ACF} & {Acc. (\%)} \\
			\midrule
			
			\checkmark  &                       &    &42.84$\pm$0.87    \\
			&  \checkmark           &     & 40.80$\pm$1.04  \\
			&                       & \checkmark & 38.48$\pm$0.09   \\
			\midrule
			
			\checkmark  &  \checkmark           &    & 44.17$\pm$3.55  \\
			\checkmark  &                       & \checkmark   & 44.60$\pm$0.43  \\
			&  \checkmark           & \checkmark  & 40.88$\pm$1.10  \\
			\midrule
			\checkmark  &  \checkmark           & \checkmark & \textbf{48.93$\pm$0.54}  \\
			
			\bottomrule
		\end{tabular}
		\label{table:diff_input_fusion}			
\end{table}

\subsubsection{Effect of Representation Combination}

We further evaluate the contribution of IQ, AP, and ACF under the AWGN$\rightarrow$Ri1 adaptation setting.
All variants use the same training strategy and fusion framework, with only the input representation combination changed. Different from Table~\ref{tab:rep_comparison}, which uses linear addition to isolate prior selection, this ablation uses the complete DKDNet training and fusion framework.
The results are summarized in Table~\ref{table:diff_input_fusion}.

Among single-representation inputs, IQ achieves the highest accuracy of $42.84\%$, confirming the importance of waveform-level information.
Adding AP or ACF to IQ further improves the accuracy to $44.17\%$ and $44.60\%$, respectively, showing their complementarity to IQ.
In contrast, AP+ACF achieves only $40.88\%$, indicating that structural cues alone are insufficient without the raw waveform observation.
The full IQ+AP+ACF combination obtains the best accuracy of $48.93\%$, outperforming all single- and two-representation variants.
These results confirm that IQ, AP, and ACF provide complementary priors: IQ preserves waveform-level discriminative information, AP provides amplitude--phase dynamics, and ACF contributes relatively stable correlation-domain statistics.
Their joint use offers the best balance among discrimination, complementarity, and domain stability for cross-domain AMC.

\begin{table}[t]
	\centering
	\caption{
		Performance of DKDNet with different AMC backbones under the AWGN$\rightarrow$Ri1 adaptation setting.
		For each backbone, Source Only denotes training only on the source domain, while Adapted denotes the corresponding DKDNet-enhanced model.
		Results are reported as mean$\pm$std over five random seeds.
	}
	\label{table:diff_backbone}
	\renewcommand{\arraystretch}{1.25}
	\setlength{\tabcolsep}{3.5pt}
	\resizebox{\columnwidth}{!}{
		\begin{tabular}{l|cc|cc}
			\toprule
			\multirow{2}{*}{Backbone}
			& \multicolumn{2}{c}{Acc. (\%) $\uparrow$ (-20 dB$\sim$18 dB)}
			& \multicolumn{2}{c}{Acc. (\%) $\uparrow$ (0 dB$\sim$18 dB)} \\
			\cmidrule(lr){2-3} \cmidrule(lr){4-5}
			& Source Only & Adapted & Source Only & Adapted \\
			\midrule
			CLDNN~\cite{7920754} 
			& 25.76$\pm$0.45 & 43.56$\pm$1.40 
			& 32.75$\pm$0.73 & 66.73$\pm$2.89 \\
			MCLDNN~\cite{xu2020spatiotemporal} 
			& 28.50$\pm$0.35 & 47.94$\pm$0.73 
			& 37.44$\pm$0.80 & 74.76$\pm$1.23 \\
			VGG~\cite{o2018over} 
			& 27.64$\pm$1.35 & 49.37$\pm$0.58 
			& 35.04$\pm$2.64 & 77.35$\pm$1.18 \\
			CNN2~\cite{liu2017deep} 
			& 27.75$\pm$0.65 & 47.84$\pm$1.26 
			& 35.64$\pm$1.16 & 74.20$\pm$2.48 \\
			LSTM~\cite{rajendran2018deep} 
			& 27.29$\pm$0.61 & {49.70$\pm$0.30} 
			& 34.94$\pm$1.35 & {77.64$\pm$0.63} \\
			DAE~\cite{ke2021real} 
			& 26.06$\pm$0.94 & 47.45$\pm$1.21 
			& 32.71$\pm$1.72 & 73.82$\pm$2.42 \\
			\bottomrule
		\end{tabular}
	}
\end{table}

\subsubsection{Flexibility across AMC Backbones}

To examine whether DKDNet depends on a specific feature extractor, we integrate it with six representative AMC backbones, including CLDNN~\cite{7920754}, MCLDNN~\cite{xu2020spatiotemporal}, VGG~\cite{o2018over}, CNN2~\cite{liu2017deep}, LSTM~\cite{rajendran2018deep}, and DAE~\cite{ke2021real}. 
For each backbone, the Source Only model is compared with its DKDNet-enhanced version under the AWGN$\rightarrow$Ri1 adaptation setting. 
As shown in Table~\ref{table:diff_backbone}, DKDNet consistently improves target-domain accuracy across all tested backbones. 
Over all SNR levels, the gains range from $17.80\%$ for CLDNN to $22.41\%$ for LSTM. 
Under high-SNR conditions, the improvements become more pronounced, ranging from $33.98\%$ to $42.70\%$. 
These results indicate that DKDNet is not restricted to a specific feature extractor and can be flexibly integrated with different AMC backbones.
Among the tested backbones, LSTM achieves the best adapted performance, reaching $49.70\%$ over all SNR levels and $77.64\%$ under high-SNR conditions, while VGG and MCLDNN also obtain competitive results. 
This suggests that DKDNet can benefit from stronger backbone designs while maintaining consistent cross-domain improvements.

\begin{table}[t]
	\centering
	\caption{
		Performance of DKDNet combined with different UDA methods under the AWGN$\rightarrow$Ri1 adaptation setting.
		The asterisk (*) denotes the DKDNet-enhanced variant of each UDA method.
		Results are reported as mean$\pm$std over five random seeds.
	}
	\label{table:compare}
	\renewcommand{\arraystretch}{1.25}
	\setlength{\tabcolsep}{4pt}
	\begin{tabular}{l|cc}
		\toprule
		{Method} 
		& \makecell[c]{{Acc. (\%) $\uparrow$}\\{(-20 dB$\sim$18 dB)}} 
		& \makecell[c]{{Acc. (\%) $\uparrow$}\\{(0 dB$\sim$18 dB)}} \\
		\midrule
		DAAMR~\cite{li2021domain} & 40.42$\pm$0.61 & 60.76$\pm$0.73 \\
		DAAMR* & 46.52$\pm$1.35 & 72.03$\pm$2.47 \\
		\midrule
		DMRUE~\cite{li2019deep} & 36.09$\pm$0.81 & 54.30$\pm$1.42 \\
		DMRUE* & \textbf{44.03$\pm$0.24} & \textbf{67.48$\pm$0.55} \\
		\midrule
		ADDA~\cite{ADDA} & 38.83$\pm$0.51 & 60.17$\pm$0.84 \\
		ADDA* & 40.01$\pm$2.62 & 62.40$\pm$5.05 \\
		\midrule
		SigDA~\cite{SigDA} & 39.32$\pm$0.16 & 61.17$\pm$0.26 \\
		SigDA* & 43.92$\pm$0.16 & 65.38$\pm$4.49 \\
		\bottomrule
	\end{tabular}
\end{table}

\subsubsection{Compatibility with UDA Methods}

To examine whether DKDNet is compatible with different domain adaptation strategies, we integrate the proposed prior-guided representation learning and fusion framework with four representative UDA methods, including DAAMR~\cite{li2021domain}, DMRUE~\cite{li2019deep}, ADDA~\cite{ADDA}, and SigDA~\cite{SigDA}. 
The enhanced variants are denoted by an asterisk (*), as shown in Table~\ref{table:compare}.
All enhanced variants outperform their original counterparts under the AWGN$\rightarrow$Ri1 adaptation setting. 
For example, DAAMR* improves the all-SNR and high-SNR accuracies by $6.10\%$ and $11.27\%$, respectively. 
DMRUE* obtains the largest gain, improving by $7.94\%$ and $13.18\%$ under the two evaluation ranges. 
SigDA* and ADDA* also achieve consistent improvements, although the gain on ADDA is relatively smaller. 
These results show that DKDNet is not tied to a specific UDA objective, but can provide more transferable features for different domain alignment strategies.

\begin{figure}[t]		
		\centering
		\includegraphics[width=0.8\linewidth]{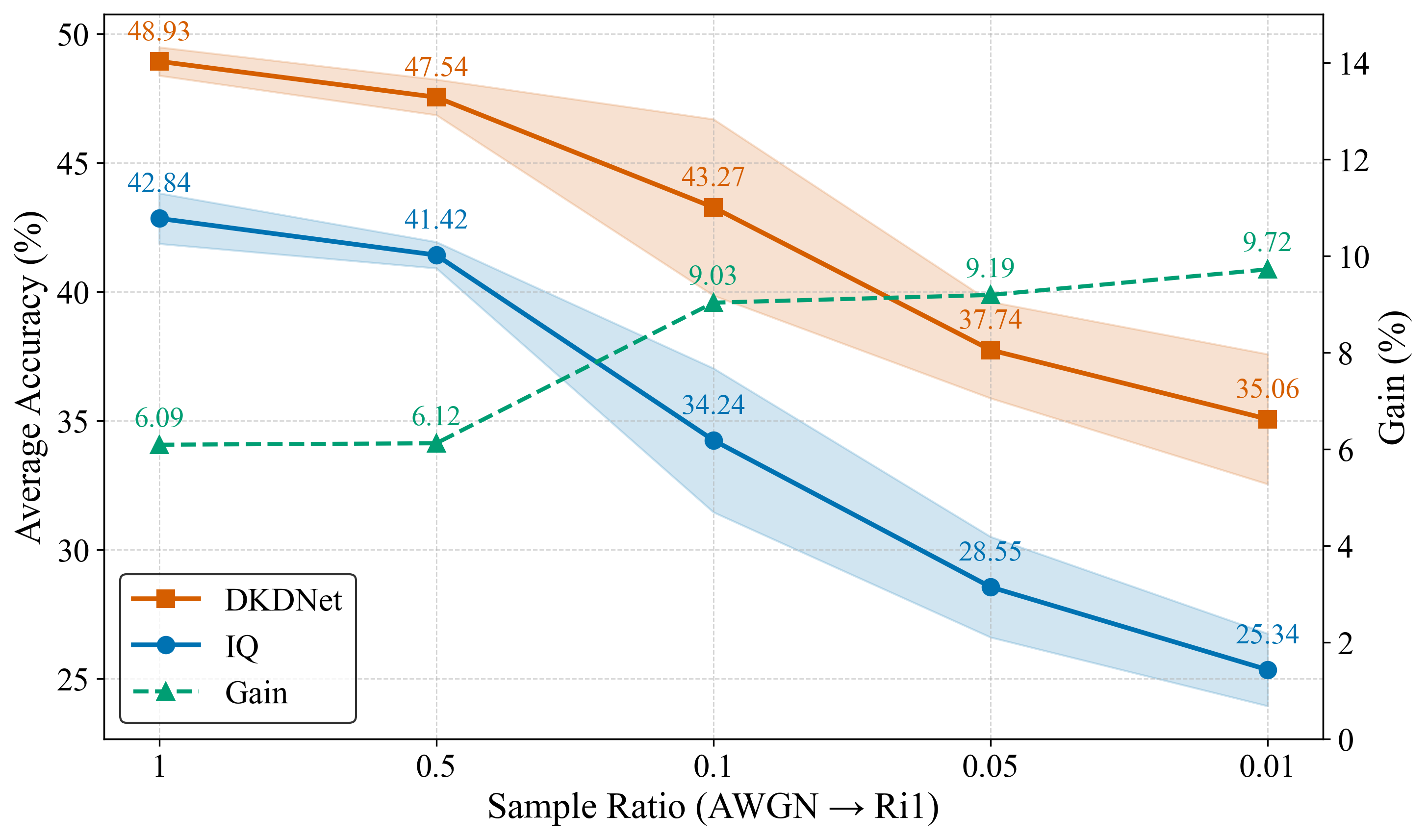}
		\caption{
			Evaluation of sample efficiency for DKDNet and the IQ-only baseline under the AWGN$\rightarrow$Ri1 adaptation setting across different training-data ratios.
		}
		\label{fig:sample}
\end{figure}

\subsubsection{Sample Efficiency under Limited Data}

To evaluate the sample efficiency of DKDNet, we compare it with a single-branch IQ-only baseline under the AWGN$\rightarrow$Ri1 adaptation setting. 
For a training-data ratio $r$, we retain the same fraction $r$ of the source-domain labeled training samples and the target-domain unlabeled training samples, while keeping the test set unchanged. 
The ratios are set to 100\%, 50\%, 10\%, 5\%, and 1\%.
As shown in Fig.~\ref{fig:sample}, DKDNet consistently outperforms the IQ-only baseline across all training-data ratios. 
Although the accuracy of both methods decreases as fewer training samples are used, DKDNet maintains a clear advantage. 
The gain increases from $6.09$ percentage points at 100\% data to $9.72$ percentage points at 1\% data. 
These results indicate that the introduced signal priors improve sample efficiency and help DKDNet retain stronger cross-domain performance when both labeled source data and unlabeled target data are limited.

\section*{Conclusions}

This paper investigates the role of signal modulation priors in improving cross-domain generalization for AMC. 
Different from existing UDA-based methods that mainly rely on data-driven feature alignment, DKDNet integrates prior-guided signal representations with data-driven feature learning. 
By analyzing representative signal representations, IQ, AP, and ACF are selected as complementary inputs to provide waveform, amplitude--phase, and correlation-domain information. 
The proposed MRFE and DLFU further enable unified representation learning and adaptive feature fusion, while adversarial domain alignment improves feature transferability across domains. 
Extensive experiments demonstrate that DKDNet effectively enhances cross-domain AMC performance, robustness, flexibility, and sample efficiency.
Despite its effectiveness, DKDNet currently focuses on a limited set of signal representations. 
In future work, more diverse and task-specific priors will be explored, and the framework will be extended to challenging scenarios such as few-shot AMC to further improve its adaptability and robustness.


\bibliographystyle{IEEEtran} 
\bibliography{reference}

\end{document}